\documentclass[twocolumn,table]{aastex62}

\usepackage{amsmath,amssymb,amstext}
\usepackage{mathtools}
\usepackage{gensymb}
\usepackage{todonotes}
\usepackage{comment}
\usepackage{graphicx}
\usepackage{multirow}
\usepackage{chngcntr}
\usepackage{booktabs}
\usepackage{xcolor}
\usepackage{soul}

\defcitealias{Barnes.Metzger_2022.ApJL_rproc.collapsar}{BM22}

\renewcommand{\thefootnote}{\fnsymbol{footnote}}

\newcommand{\nickel}{\ensuremath{^{56}{\rm Ni}}}
\newcommand{\mej}{\ensuremath{M_{\rm ej}}}
\newcommand{\ekin}{\ensuremath{E_{\rm k}}}
\newcommand{\mni}{\ensuremath{M_{56}}}

\newcommand{\msun}{\ensuremath{M_{\odot}}}

\newcommand{\per}[1]{#1\ensuremath{^{-1}}}
\newcommand{\Mw}{\ensuremath{M_{\rm w}}}
\newcommand{\tauw}{\ensuremath{\tau_{\rm w}}}
\newcommand{\tiw}{\ensuremath{t_{\rm 0,w}}}
\newcommand{\vw}{\ensuremath{v_{\rm w}}}
\newcommand{\Ew}{\ensuremath{E_{\rm w}}}
\newcommand{\Xw}{\ensuremath{X_{\rm w}}}
\newcommand{\mbar}{\ensuremath{\overline{m}}}
\newcommand{\Ei}{\ensuremath{E_0}}

\newcommand{\rp}{\emph{r}-process}
\newcommand{\rccsne}{\emph{r}CCSNe}
\newcommand{\rccsn}{\emph{r}CCSN}
\newcommand{\crp}{collapsa\emph{R}-process}

\shorttitle{Collapsar wind mixing} 
\shortauthors{Barnes \& Duffell}

\begin{document}

\title{Hydrodynamic mixing of accretion disk outflows in collapsars: implications for \rp{} signatures}

\author[0000-0003-3340-4784]{Jennifer Barnes}
\affil{Kavli Institute for Theoretical Physics, Kohn Hall, University of California, Santa Barbara, CA 93106, USA }
\email{jlbarnes@kitp.ucsb.edu}
\author[0000-0001-7626-9629]{Paul C. Duffell}
\affiliation{Department of Physics and Astronomy, Purdue University, 525 Northwestern Avenue, West Lafayette, IN 47907, USA}

\begin{abstract}
The astrophysical environments capable of triggering heavy-element synthesis via rapid neutron capture (the \rp) remain uncertain.
While binary neutron star mergers (NSMs) are known to forge \rp{} elements, certain rare supernovae (SNe) have been theorized to supplement---or even dominate---\emph{r}-production by NSMs.
However, the most direct evidence for such SNe, unusual reddening of the emission caused by the high opacities of \rp{} elements, has not been observed.
Recent work identified the distribution of \rp{} material within the SN ejecta as a key predictor of the ease with which signals associated with \rp{} enrichment could be discerned.
Though this distribution results from hydrodynamic processes at play during the SN explosion, thus far it has been treated only in a parameterized way.
We use hydrodynamic simulations to model how disk winds---the alleged locus of \emph{r}-production in rare SNe---mix with initially \rp-free ejecta.
We study mixing as a function of the wind mass and duration and of the initial SN explosion energy, and find that it increases with the first two of these and decreases with the third.
This suggests that SNe accompanying the longest long-duration gamma-ray bursts are promising places to search for signs of \rp{} enrichment.
We use semianalytic radiation transport to connect hydrodynamics to electromagnetic observables, allowing us to assess the mixing level at which the presence of \rp{} material can be diagnosed from SN light curves.
Analytic arguments constructed atop this foundation imply that a wind-driven \rp-enriched SN model is unlikely to explain standard energetic SNe.
\end{abstract}

\section{Introduction} \label{sec:intro}

The high free-neutron flux required for the synthesis of heavy elements through rapid neutron capture \citep[the \rp;][]{Burbidge2.Howler.Foyle_1957.RMP_rprocess, Cameron_1957.AJ_rprocess} has rendered the astrophysical site(s) of \rp{} production an enduring mystery.
The multimessenger detection of the neutron star merger (NSM) GW170817 \citep{Abbott.ea_2017ApJL_gw.170817.multimess} confirmed \citep{Drout.ea_2017Sci_gw.170817.emcp.disc,Kasen.ea_2017Natur_gw.170817.knova.theory,Kasliwal.ea_2017Sci_gw.170817.em.interp,Tanaka.ea_2017PASJ_gw.170817.knova.interp,Watson.ea_2019.Natur_gw170817.knova.strontium} the long-standing theory that the decompression of NS material following its disruption during a merger could trigger an \rp{} \citep{Lattimer.Schramm_1974.ApJL_rproc.nsbh.merger,Lattimer_1976,Symbalisty.Schramm_1982.ApL_rproc.ns.merger,Meyer_1989.ApJ_rproc.nsm,Davies.ea_1994.ApJ_nsm.model,Freiburghaus.ea_1999.ApJL_rproc.nsm}.
However, questions remain as to whether NSMs can explain the full pattern of \rp{} enrichment observed across time and space \citep[e.g.,][]{Cote.ea_2019.ApJ_other.rproc.sources,Zevin.ea_2019.ApJ_nsm.rproc.glob.cluster,vandeVoort.ea_2020_rproc.enrich.nsm.rare.ccsne,Jeon.ea_2021.mnras_rproc.stars.ufdwarf.gal,Molero.ea_2021.mnras_evol.rproc.dwarf.gal,delosReyes.ea_2022.apj_sfh.nucleosyn.sculptor,Naidu.ea_2022.ApJ_disrupted.halos.rproc.sources,Cavallo.ea_2023.arXiv_Eu.hierarchical.gal.form,Kobayashi.ea_2023.ApJ_nsm.rproc.milkyway}.

Amid these uncertainties, ``collapsars''---the core-collapse supernova (CCSN) explosions of rapidly rotating massive stars \citep{MacFadyenWoosley99_collapsar}---have been investigated as a possible \rp{} production site \citep{Pruet.ea_2003.ApJ_nucleosyn.grb.disks,Surman.McLaughlin_2004.ApJ_neutrinos.nucleo.grb.disks,Surman_2006}.
While conditions in the accretion disks that form post-collapse were found to support neutronization in the disk midplane, it was not clear that this material could remain neutron rich in the face of successive neutrino absorptions, which were believed to be the mechanism responsible for ejecting matter from the disk.
Recently, \citet{Siegel.Barnes.Metzger_2019.Nature_rp.collapsar} proposed that magnetic turbulence unbinds the newly neutron-rich  material, limiting opportunities for neutrino absorption and allowing the resulting wind to undergo an \rp{} as it expands.
However, simulations of collapsar disks employing different neutrino transport schemes disagree on how neutron-rich any ejected material would be, and therefore on the plausibility of collapsars as sites of robust heavy element production \citep{Miller.ea_2020.ApJ_rproc.collapsar.blue,Fujibayashi.ea_2022.arXiv_collapsars,Just.ea_2022.ApJ_rproce.collapsar.disk}.
Avenues beyond simulation may help break this impasse.

The collapsar model was originally conceived of \citep{MacFadyenWoosley99_collapsar} as an explanation for long gamma-ray bursts (GRBs) and the unusually energetic ``broad-lined'' Type Ic supernovae (SNe Ic-BL) observed to accompany them \citep{Galama98_bwDisc,Iwamoto.ea_1998.Natur_icbl.model.1998bw,Woosley.ea_1999.ApJ_icbl.model.1998bw,Mazzali.ea_2003.ApjL_icbl.sn.2003dh}.
The \rp{} collapsar hypothesis exists within this framework. 
If collapsars eject \rp-rich disk winds, these winds will be embedded in the SN ejecta, and the uniquely high opacity of heavy \rp{} compositions \citep{Kasen_2013_AS,Tanaka_Hotok_rpOps,Tanaka.ea_2020.MNRAS_knova.kappas} could redden the SN emission relative to what would be expected for an \rp-free explosion.
This points to SN observations as an important tool for assessing \emph{r}-production in collapsars, i.e., the \crp.

The potential impact of \rp{} material on SN signals was mentioned by \citet{Siegel.Barnes.Metzger_2019.Nature_rp.collapsar}, but was systematically studied only later by \citet[][hereafter \citetalias{Barnes.Metzger_2022.ApJL_rproc.collapsar}]{Barnes.Metzger_2022.ApJL_rproc.collapsar}, who used semianalytic radiation transport methods to model emission from \rp-enriched CCSNe (\rccsne) across a broad sector of parameter space.
They found that the extent of the reddening depends sensitively on how thoroughly the \rp{} elements are mixed into the ordinary SN ejecta, with the degree of mixing a free parameter of their model. 

If an \rccsn{} launches a wind, some mixing is expected generically, due to hydrodynamic instabilities thought develop at the interface between the high-velocity wind and the presumably slower ejecta composed of ordinary stellar material and/or explosively synthesized \nickel.
However, a complete understanding of mixing requires knowledge about the nature of the explosions that give rise to GRBs and SNe Ic-BL, which remains a topic of active inquiry \citep[e.g.,][]{Burrows.ea_2007_sn.rotation,Kumar.ea_2008.MNRAS_grb.central.engine.fallback,Mosta.ea_2015Natur_magnetorotational.ccsne,Sobacchi17a_SNjetEngine,Gottlieb.ea_2022.mnras_bh_3d.grmhd.ccsn.jets,Gottlieb.ea_2022.ApJL_bh.to.photosphere,Eisenberg.ea_2022.MNRAS_icbl.two.component.expls,Halevi.ea_2023.ApJ_rho.profs.collapsars}.

The large quantities of \nickel{} inferred for SNe Ic-BL \citep{Prentice.ea_2016_SeSne.bolLCs,Taddia.ea_2019.AandA_IcBL.iPTF.survey} have often, in the context of the collapsar model, been attributed to \nickel-burning in collapsar disk outflows \citep[e.g.,][]{Pruet.ea_2003.ApJ_nucleosyn.grb.disks,Nagataki.ea_2006.ApJ_hypernova.nucleosyn,Surman_2006}.
However, the \rp{} collapsar scenario holds that these outflows instead burn heavier elements, and so requires a distinct mechanism to explain observed SNe Ic-BL \nickel{} masses.
A favored---though not universal---alternative to the \nickel-wind scenario is a prompt explosion phase that rapidly injects energy into the inner layers of the collapsing star \citep{Maeda.Tominaga_2009.MNRAS_nickel.wind.driven.sne,Suwa.Tominaga_2015.MNRAS_ni56.magnatar.hypernovae}.

The general picture of the \crp{} thus includes a prompt explosion, the subsequent formation and dissipation of an accretion disk, and, in some cases, an ultrarelativistic GRB jet.
Each of these processes has the potential to influence the dynamics of the SN explosion, but the manner in which they fit together---their relative importance and even their chronology---is uncertain, motivating a survey of mixing behavior over a wide range of explosion models.

We perform hydrodynamic calculations of wind-ejecta mixing in collapsar-generated SN outflows.
Our hydrodynamics set-up is described in \S\ref{sec:methods}, and the results of our calculations can be found in \S\ref{sec:hydro-results}.
In \S\ref{sec:sn_emission}, we use radiation transport to predict light curves for a subset of our  models, and discuss the implications of our results for efforts to observe \rccsne.
We contextualize our findings and discuss future research directions in \S\ref{sec:conclusion}.

\section{Methods}\label{sec:methods}

We model collapsar wind mixing with the special relativistic, moving-mesh hydrodynamics code \texttt{Jet} \citep{Duffell_2013_JET,Duffell15_jetInject}, which we have (ironically) adapted to simulate a spherical wind outflow. 

\subsection{Stellar Progenitor}\label{subsec:methods-prog}

We begin with an analytic progenitor model representative of the stripped-envelope stars generally presumed \citep[e.g.,][]{Yoon.Langer_2005.AandA_grb.progenitors,Woosley.Heger_2006.ApJ_grb.progenitors,Modjaz.ea_2014_se-sn.dat,Liu.Modjaz.Bianco_2016.ApJ_Hpoor.SlSne,Taddia.ea_2019.AandA_IcBL.iPTF.survey} to explode as GRB-SNe and SNe Ic-BL.
We used the same progenitor in earlier work on jet-driven SNe \citep{Barnes.Duffell.ea_2018.ApJ_grb.icbl.engine}.
It corresponds to an evolved, stripped star with a pre-collapse mass and radius of $5.0\msun$ and $1.6 R_\odot$, respectively.

We assume that the innermost layers of the progenitor have collapsed to a  black hole, and approximate the effects of this collapse by introducing a low-density cavity interior to $r_{\rm cav} = 9\times 10^{-4} R_0$, with $R_0$ the progenitor radius.
The progenitor is spherically symmetric and has a mass density that depends on the radial coordinate as 
\begin{align}
    \rho_0(r) = \: &\alpha_\rho \; \frac{0.0615 M_0}{R_0^3} \left(\frac{r}{R_0}\right)^{-2.65}\left( 1- \frac{r}{R_0}\right)^{3.5}, \label{eq:rho_prog}\\\
    \text{where } \:\:\: &\alpha_\rho = \nonumber
    \begin{cases}
    1 &\hfill \text{ for } r \geq r_{\rm cav} \\
    10^{-3} &\hfill \text{ for } r < r_{\rm cav},
    \end{cases}
\end{align}
and $M_0$ is the mass outside the cavity.
To ensure numerical tractability, we set the density within the cavity according to Eq.~\ref{eq:rho_prog}.
However, we do not resolve the central remnant or the surrounding accretion disk.
The pre-collapse progenitor mass is therefore greater than the total mass on our computational grid.

\subsection{Hydrodynamics set-up}\label{subsec:methods-hydro}
Our initially stationary progenitor is exploded by an accretion disk wind and, in some cases, an additional prompt explosion.
We parameterize the disk wind by its mass, \Mw, its characteristic timescale \tauw, its velocity \vw, and the time at which it begins, \tiw.
The wind is injected into the progenitor via source terms in energy,
\begin{gather}
    S_{\rm E} = \dot{E}_{\rm w} \times \frac{r}{8 \pi r_0^4} \exp[-r^2/2r_0^2] \; f(t), \label{eq:source_en}\\
    \text{with } \:\: f(t) = \begin{cases} \exp\left[-(t-\tiw)/\tauw\right] \:\: &\text{for } t\geq \tiw\nonumber \\
    0 & \text{for } t < t_{0,\rm w}; \nonumber
    \end{cases}
\shortintertext{mass,}
S_{\rm M} = \frac{S_{\rm E}}{\vw^2/2};\label{eq:source_mass} \\
\shortintertext{and radial momentum,}
S_{\rm P,r} = \vw S_{\rm E}\label{eq:source_mom}.
\end{gather}
In Eq. \ref{eq:source_en}, $r_0 = 2 \times 10^{-4} R_0 < r_{\rm cav}$ is the radius at which the wind injection peaks, and $\dot{E}_{\rm w} = \Mw \vw^2/2\tauw$ is the characteristic wind power.
Eqs. \ref{eq:source_en}--\ref{eq:source_mom} ensure that the time- and volume-integrated mass (energy) introduced into the progenitor is \Mw{} ($\Mw \vw^2/2$).
\footnote[1]{We note that our formulation for radial momentum (Eq.~\ref{eq:source_mom}) is more traditionally applied to highly relativistic systems. 
For a more modestly relativistic flow, like our wind, the effect is to seed the outflowing material with a combination of kinetic and thermal energy. 
Because we inject the wind within a cavity, its expansion in a low-density medium enables the efficient conversion of thermal to kinetic energy, and when it reaches the cavity boundary it has accelerated to ${\sim}\vw$.
}

We track the distribution of wind material in the ejecta from \tiw{} until the end of the simulation using a passive scalar.

In the parlance of this project, models that also undergo a prompt explosion are said to have a non-zero initial explosion energy \Ei.  
We induce a prompt explosion in these models by seeding the inner layers of the progenitor with excess thermal energy, which takes the form of an additional pressure term,
\begin{equation}
    P_{\rm exp}(r) = \frac{(\Gamma - 1)\Ei}{\pi^{3/2} ~r_{\rm exp}^3} \times \exp[-(r/r_{\rm exp})^2],
\end{equation}
where $\Gamma = 4/3$ is the adiabatic index, $E_0$ is the energy of the prompt explosion, and $r_{\rm exp} = 2r_{\rm cav}$.
The total pressure in the progenitor, prior the introduction of the wind, is 
\begin{align*}
    P(r) = \begin{cases}
    P_{\rm exp}(r) + 10^{-6}\rho(r) \:\: &\text{ for } E_0 > 0 \\
    \min\{ 10^{-6}\rho(r),\; M_0 R_0^{-3} \} \:\: &\text{ for } E_0 = 0,
    \end{cases}
\end{align*}
with the behavior for $\Ei = 0$ approximating the limit of a cold gas.

At least some of the collapsars believed to generate \rp{} disk outflows are likely to be accompanied by ultrarelativistic GRB jets. (Indeed, the analogy to the short GRBs tied to NSMs was the foundation upon which \citet{Siegel.Barnes.Metzger_2019.Nature_rp.collapsar} argued for the \crp.)
Nevertheless, for the sake of simplicity we consider here only the wind and the prompt explosion. 
We suspect based on earlier work \citep{Barnes.Duffell.ea_2018.ApJ_grb.icbl.engine} that a jet could increase mixing, but that its impact would be restricted to the narrow region just beyond the jet opening angle. 
The calculations here therefore provide a conservative lower limit on mixing. 

We use \texttt{Jet} to evolve the SN-wind system on a two dimensional, axisymmetric grid divided into zones of size $\delta r/r \sim 4.5 \times 10^{-5}$ and $\delta \theta \sim 6 \times 10^{-3}$.
Our calculations begin at a time 1000 times less than the light-crossing time of the progenitor (i.e., $t_0 = 10^{-3} R_0/c$) and continue until $t \gtrsim 1000\times R_0/c$.
By this point, the ejecta have expanded to a few hundred times the original radius of the progenitor, and the flow has become homologous.
As a result, the hydrodynamic instabilities that drive mixing have frozen out, and the distribution of wind matter within the progenitor has achieved its final state.

\subsection{Description of Models}\label{subsec:methods-models}
Our model suite, which we summarize in Table~\ref{tab:params}, explores the impact of three of our four wind parameters (we fix $\vw = 0.1c$) and \Ei{} on wind-ejecta mixing.
We vary the wind mass, \Mw, within the range $0.01 \leq \Mw/M_0 \leq 0.5$.

To explore the role of the prompt explosion,
we select three values of $E_0$ corresponding to average kinetic velocities ($v_{\rm char,0} = (2E_0/M_0)^{1/2}$ of $0.01c$, $0.03c$, and $0.1c$.
In other words, the chosen $E_0$ reflect inferred velocities of fairly slow, typical, and fast (broad-lined) SNe Ic. 
Wind-driven models have no prompt explosion ($\Ei=0$), and derive their kinetic energy exclusively from disk outflows.  

We also examine the effect of the wind duration, \tauw.
We adopt as a fiducial case $\tauw = R_0/c$, but for certain combinations of \Mw{} and $E_0$, we consider durations that differ from the fiducial value by a factor of up to 10 in either direction.
(See the beginning of \S\ref{subsec:tau_wind} for a discussion of our choice of wind durations.)

Finally, we explore how the wind start, \tiw, impacts mixing.
In prompt-explosion models, we generally assume that the wind and the explosion begin simultaneously when the simulation starts, at $t_0 = 10^{-3}R_0/c$.
However, disk outflows could lag the prompt explosion if it takes time for the disk to form and accretion to begin \citep[e.g.,][]{Kohri.ea_2005.ApJ_disk.wind.sne}.
Thus, we also consider ``delayed-wind'' models, for which \tiw{} ranges from (0.1--10)$R_0/c$.
For wind-driven models, the progenitor is static until the wind begins, so altering \tiw{} has no effect on the outcome of the simulation.

As mentioned above, one variable we omit from the current study is \vw, which is $0.1c$ in all cases.
This choice was motivated by studies of disk outflows, which predict wind velocities narrowly clustered around the disk's escape velocity $v_{\rm esc} \approx 0.1c$ \citep[e.g.,][]{MacFadyenWoosley99_collapsar}.
However, for certain magnetic field configurations, $\vw$ could reach 0.2--0.3$c$ \citep{Christie.ea_2019.MNRAS_bfield.geom.nsm.discs}.
We briefly revisit the question of $\vw$ in \S\ref{subsec:windy-sn}.

\begin{table}[h]
\begingroup
\centering
\caption{Parameters of the model suite}\label{tab:params}
    \begin{tabular}{>{\raggedright\arraybackslash}p{0.17\columnwidth}p{0.36\columnwidth}>{\raggedleft\arraybackslash}p{0.36\columnwidth}}
    \toprule
     \emph{Parameter} & \emph{Values}  & \emph{Notes} \tabularnewline
    \hline
    $\Mw$ [$M_0$] & 0.01, 0.05, 0.1, 0.2, 0.3, 0.5 & --- \\
    $E_0$ [$M_0 c^2$] & 0, $5\times 10^{-5}$, $4.5 \times 10^{-4}$, $5\times 10^{-3}$  & with no wind, $\{E_0>0\}$ would yield average velocities of 0.01$c$, 0.03$c$, \& 0.1$c$. \\
    \tauw{} [$R_0/c$] &
     0.1, 0.3, 1.0, 3.0, 10.0 & $\neq 1.0$ only for $\Mw = 0.1$, $E_0 \in \{0, 4.5 \times 10^{-4}\}$ \\
    $\tiw \; [R_0/c]$ & $10^{-3}$, 0.1, 0.3, 1.0, 3.0 10.0 & ${>}10^{-3}$ only for $\Mw = 0.1$, $E_0 = 4.5 \times 10^{-4}$ \\
    $\vw \; [c]$ & 0.1 & --- \\
    \bottomrule
    \end{tabular}
    \endgroup
\end{table}

\section{Hydrodynamics Results}\label{sec:hydro-results}

Our hydrodynamic calculations predict the final density profile of the ejecta and the distribution of the \rp{} wind matter within it. 

\subsection{Mixing metrics}\label{subsec:hydro-metrics}
The combination of the disk wind and, if present, the prompt explosion accelerates the ejecta to velocities $v \sim 0.01$--$0.1c$.
The wind inflates a lower-density bubble in the center of the ejecta, sweeping the ejected material into a shell whose velocity coordinate scales with the square root of the outflow's final kinetic energy ($v_{\rho} \propto (E_0 + \Mw\vw^2/2)^{1/2}$).

Material from the wind is concentrated behind the density peak.
In velocity space, the wind mass fraction (\Xw) is effectively a step function that transitions from $\Xw=1$ to $\Xw=0$ at $v=v_{\rho}$.
However, since the peak contains a large fraction of the system's total mass, the distribution of wind matter with respect to mass coordinate (a formulation we favor to facilitate comparison with \citetalias{Barnes.Metzger_2022.ApJL_rproc.collapsar}) is often broader.

\begin{figure}\includegraphics[width=\columnwidth]{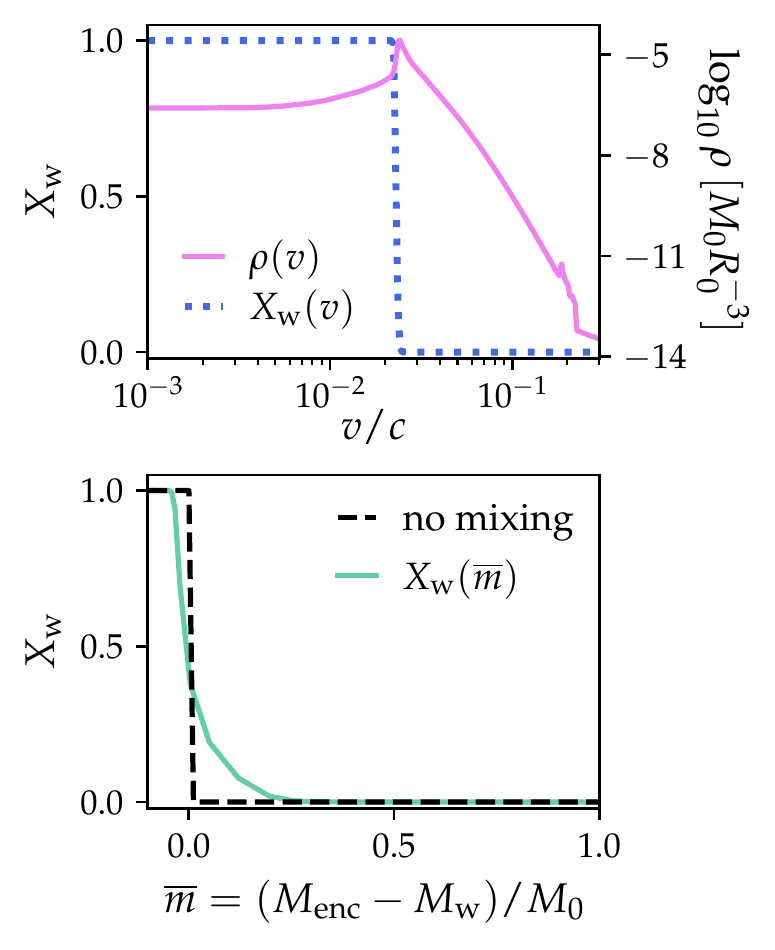}
\caption{The disk wind sweeps mass into a narrow shell, producing a spike in the density profile. 
Behind the spike, the composition is nearly purely wind; there is limited mixing of wind matter beyond the peak.
\emph{Top panel:} The final mass density $\rho$ and wind mass fraction \Xw{} as a function of velocity. 
\emph{Bottom panel:} \Xw{} as a function of the modified interior mass coordinate, $\mbar = (M_{\rm enc}-\Mw)/M_0$, where $M_{\rm enc}$ is the enclosed mass.
The dashed black curve shows \Xw{} in the limit of no mixing.
The results here are for a prompt-explosion model with $\Mw/M_0 = 0.1$, $E_0/M_0c^2 = 5\times 10^{-5}$, $\tauw=R_0/c$, and $\tiw = 10^{-3}R_0/c$. 
However, the behavior they illustrate is representative of models throughout the parameter space.
}\label{fig:wind_sn_example}
\end{figure}

To illustrate these concepts, we show in Fig.~\ref{fig:wind_sn_example} the final angle-averaged mass density profile and wind mass fraction as a function of velocity for a prompt-explosion model with $\Mw/M_0 = 0.1$, $E_0/M_0c^2 = 5 \times 10^{-5}$, and default time parameters $\tauw = R_0/c$ and $\tiw = 10^{-3}R_0/c$.
The bottom panel shows how \Xw{} varies with the modified mass coordinate $\mbar = (M_{\rm enc}-\Mw)/M_0$, with $M_{\rm enc}$ the enclosed mass.
We have devised $\mbar$ to simplify the comparison of mixing in models with different \Mw.

\subsection{Survey of the \Mw-$E_0$ landscape}\label{subsec:mw_ew}

We first investigate the effects on mixing of \Mw{} and $E_0$, while holding constant the wind duration ($\tauw=R_0/c$) and, for prompt-explosion models, the wind start time ($\tiw = 10^{-3}R_0/c$, coincident with the start time of the simulation).

\begin{figure}\includegraphics[width=\columnwidth]{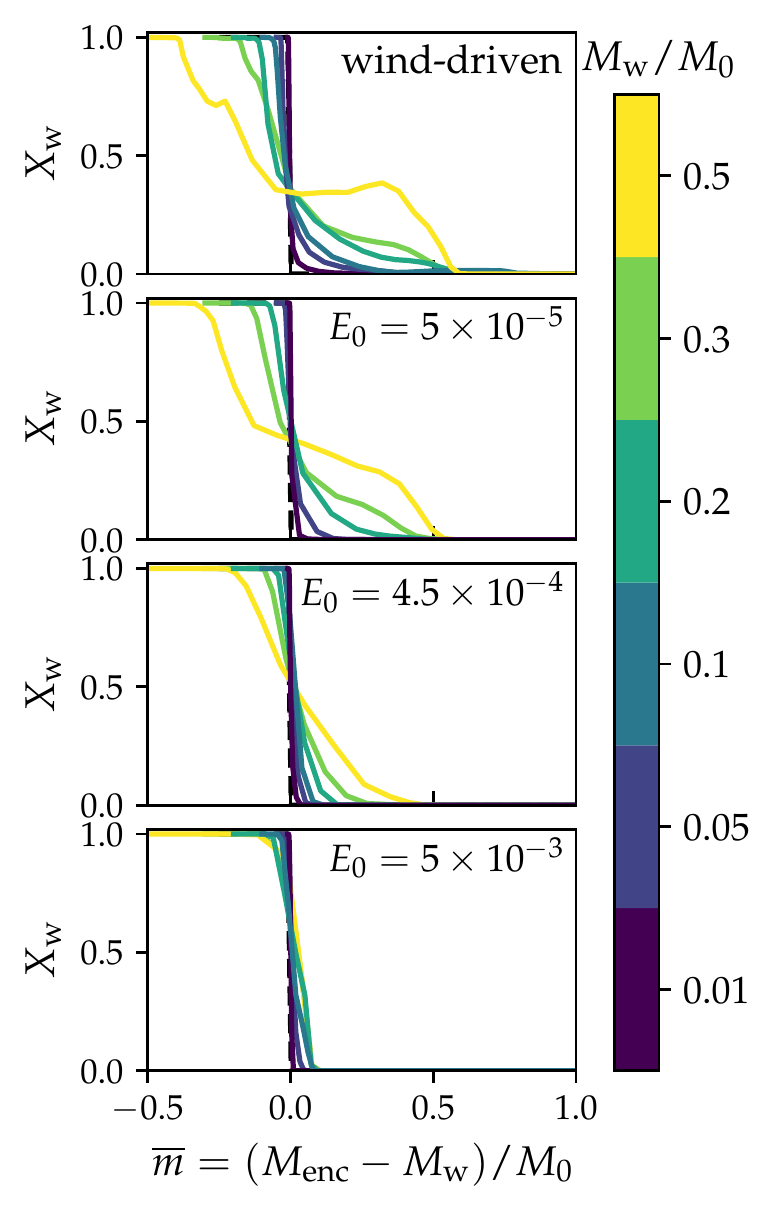}
    \caption{The mixing of the wind material into the non-wind ejecta increases with the wind mass and decreases with the energy of the prompt explosion.
    Each panel shows the wind mass fraction \Xw{} as a function of the modified mass coordinate \mbar{} for several wind masses $\Mw$ and a single prompt-explosion energy $E_0$, which we record in units of $M_0 c^2$. For wind-driven models (\emph{top panel}), $\Ei = 0$.
    While increasing \Mw{} results in more mixing regardless of \Ei, the impact of \Mw{} declines as the prompt explosion becomes more energetic.
    If \Ei{} is large enough to accelerate the ejecta to velocities comparable to \vw{} (which is itself comparable to the velocities inferred for the most energetic SNe Ic-BL), mixing is minimal regardless of \Mw{} (\emph{bottom panel}).
    }
    \label{fig:xmix_ME}
\end{figure}

The combination of \Mw{} and \Ei{} affects both the density profile of the resulting outflow and the level of mixing. 
The total kinetic energy of the explosion-plus-wind system is $\ekin = \Ei + \Mw\vw^2/2$, and increasing either term on the R.H.S. produces a flatter density spike (e.g., Fig.~\ref{fig:wind_sn_example}) at a higher velocity coordinate.
This allows the wind to mix to out to higher velocities, but, because the entire outflow expands faster, that does not necessarily correspond to greater $\mbar$.
We find that mixing is stronger for larger wind masses, but the effect of \Mw{}---as well as the overall level of mixing---diminishes with $E_0$, as shown in Fig.~\ref{fig:xmix_ME}.

The Rayleigh-Taylor instabilities that allow the wind to mix into the initially wind-free ejecta are driven by the acceleration of the slower outer material by the faster wind emanating from the core.
The wind's ability to accelerate this outer ejecta depends on the energy it carries, as well as on the initial velocity difference between the wind and the matter ejected during the prompt explosion.

We find that the dependence of  mixing on \Mw{} and $E_0$ for these models can be captured by a parameter, which we call $\zeta$, proportional  to the increase in the product of the ejecta's total mass and average velocity due to the addition of the wind,

\begin{align}
    \zeta(\Mw, \Ei) &= \left[2\left(1 + \frac{\Mw}{M_0}\right)\left(\frac{\Ei + E_{\rm w}}{M_0 c^2}\right)\right]^{1/2} \nonumber \\
    &\:\:\:\:\:\; - \left[\frac{2\Ei}{M_0 c^2}\right]^{1/2}, \text{ where} \label{eq:zeta_ME} \\
    \Ew &= \Mw\vw^2/2 \nonumber
\end{align}
is the kinetic energy of the wind, which (since \vw{} does not vary) is a function solely of \Mw, and we treat the mass of the non-wind material, $M_0$ as fixed.
In Fig.~\ref{fig:mix_by_zeta}, we plot \Xw{{} for all the models of Fig.~\ref{fig:xmix_ME}, color-coded by $\zeta$ to demonstrate the validity of this parameterization.

Figs.~\ref{fig:xmix_ME} and \ref{fig:mix_by_zeta} make clear that the level of mixing is not a function only of the wind properties, but instead depends on the interplay between the wind and any additional engine supplying energy to the ejecta.
Simulations \citep[e.g.,][]{Miller.ea_2020.ApJ_rproc.collapsar.blue,Just.ea_2022.mnras_nuetrino.cooled.bh.acc.disks} predict that accretion disks will launch winds at velocities of ${\sim}0.05$--$0.1c$, similar to the velocities that characterize the energetic SNe Ic-BL---including GRB-SNe---most likely to harbor such disks.
However, if these high-velocity SNe acquire most of their kinetic energy in a prompt explosion, even such high wind velocities will not lead to extensive mixing.
On the other hand, if very kinetic SNe have the high velocities they do \emph{because} they are accelerated by a wind, more mixing is expected.

 \begin{figure}\includegraphics[width=\columnwidth]{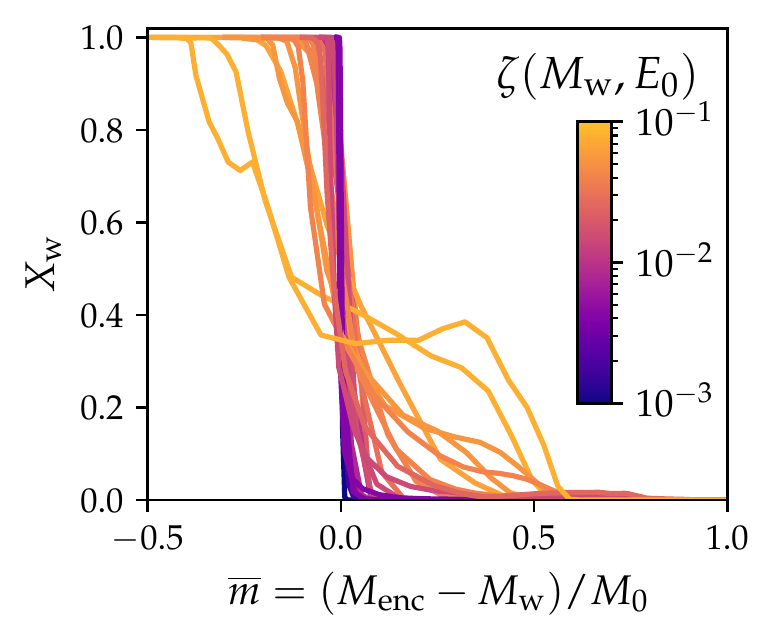}
\caption{The mixing of the wind into the wind-free ejecta increases with both the wind mass \Mw, and the importance of the wind as a source of kinetic energy for the SN.
We show above the wind mixing profiles for the models of Fig.~\ref{fig:xmix_ME}, color-coded by the parameter $\zeta$. 
This quantity, which we define in Eq.~\ref{eq:zeta_ME}, provides a handle on the additional mass and velocity the wind imparts to the explosion.
}
\label{fig:mix_by_zeta}
\end{figure}

\subsection{Effect of wind duration}\label{subsec:tau_wind}

The models of \S\ref{subsec:mw_ew} share a single wind duration, equal to the light-crossing time of the star ($\tauw = R_0/c$).
In nature, the duration of the wind is related to the dissipation time of the accretion disk that supplies it.
Since accretion disks are also the presumed engines of the relativistic jets that give rise to long GRBs \citep[e.g.,][]{MacFadyenWoosley99_collapsar,Narayan.ea_2001.ApJ_accretion.grbs}, GRB durations should reflect disk---and therefore wind---lifetimes. 
However, this correspondence is likely to be most robust when engine, and ergo GRB, timescales are much longer than the time the jet requires to break out of the progenitor star \citep[or the SN ejecta; e.g.,][]{DeColle.ea_2022.MNRAS_grb.sne.coccoon.landscape}.
In the case of shorter engines, hydrodynamic processes within the jet-shocked ``coccoon'' can extend the the jetted, relativistic flow and therefore the GRB lifetime, which may obscure the relationship between engine and GRB durations \citep{Barnes.Duffell.ea_2018.ApJ_grb.icbl.engine}.
Thus, while the observed timescales of GRBs classified as ``long'' \citep[2 s $\lesssim T_{90} \lesssim$ 100 s;][]{Kouveliotou.ea_1993.ApJ_grb.two.classes} on their own imply wind lifetimes that span nearly two orders of magnitude, the lower bound on this range is conservative, and the actual variance could be greater.

The absence of hydrogen and helium in the spectra of GRB-SNe and SNe Ic-BL more generally \citep{Modjaz.ea_2014_se-sn.dat,Liu.Modjaz.Bianco_2016.ApJ_Hpoor.SlSne} implies that the progenitor star has undergone significant stripping.
Fully stripped stars exhibit less variation in their pre-explosion radii than their partially stripped counterparts, which can expand dramatically before collapse as their hydrogen and/or helium envelopes swell 
\citep{Laplace.ea_2020.A&A_expansion.stripped.stars}.
Nevertheless, models of the evolution of SNe Ic-BL progenitors \citep{Aguilera.Dena.ea_2018.ApJ_icbl.progenitors} suggest that the terminal radii of these stars range from ${\sim}0.5 R_\odot$ to ${\gtrsim}3 R_\odot$.
If we assume no correlation between engine duration and $R_0$, the variation in each of these parameters points to an engine/wind timescale of $0.3R_0/c \lesssim \tauw \lesssim 100R_0/c$, with $T_{90} = 2 \text{ s and } R_0 = 3.0R_\odot$ ($T_{90} = 100 \text{ s and } R_0 = 0.5 R_\odot$) defining the lower (upper) bound.

To probe the effect of this variability, we vary \tauw{} from $0.1R_0/c$ to $10R_0/c$ for models with $\Mw = 0.1M_0$ and initial explosion energies $E_0/M_0c^2 = 0$, $4.5 \times 10^{-4}$, and $5\times 10^{-3}$.
For $\Ei > 0$, the wind start time \tiw{} is coincident with the start time of the simulation.
We present the final mixing profiles in Fig.~\ref{fig:tau_fx}.

\begin{figure}\includegraphics[width=\columnwidth]{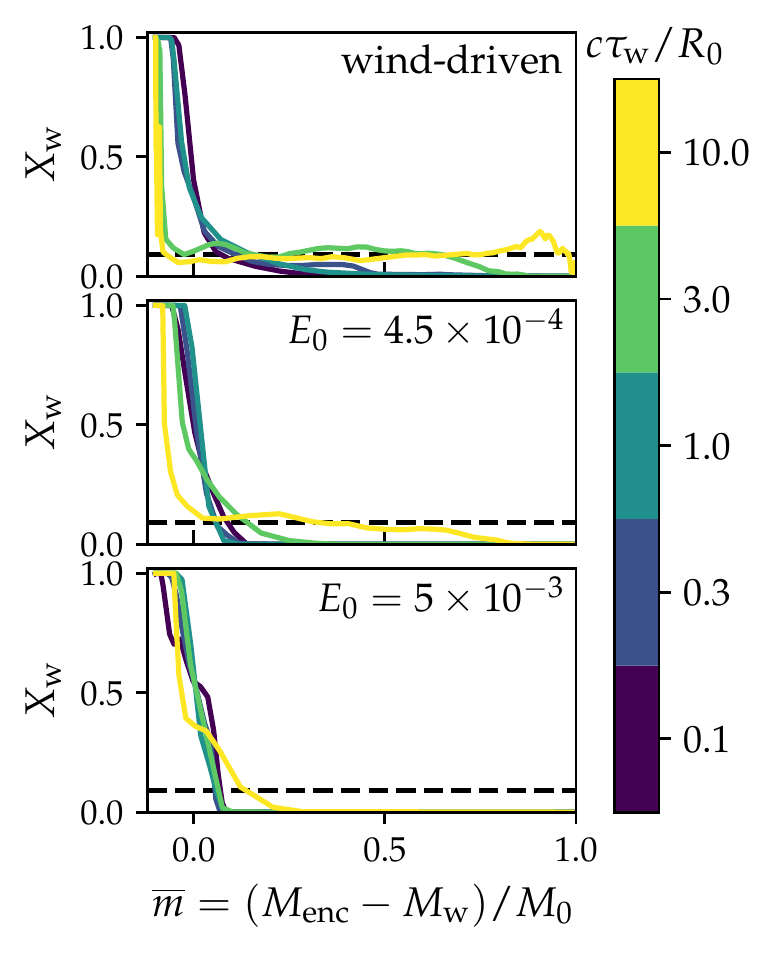}
    \caption{The level of mixing increases with the duration of the wind \tauw{}, but the sensitivity of the mixing profile to \tauw{} decreases as the energy of the prompt explosion rises.
    Above, we show the wind mass fraction $\Xw$ as a function of the modified mass coordinate \mbar{} for models with $\Mw=0.1M_0$, various prompt explosions energies \Ei{} (normalized to $M_0c^2$), and $0.1R_0/c \leq \tauw \leq 10R_0/c$.
    For the lowest \Ei{} and highest \tauw{} values considered (i.e., $0$ and $10R_0/c$ respectively; yellow curve, top panel), the wind material is close to uniformly distributed in the ejecta.
    In all panels, dashed black horizontal lines indicate \Xw{} for the case of perfect mixing.}
    \label{fig:tau_fx}
\end{figure}

Mixing increases with \tauw{}. 
In some cases, the wind profile is nearly uniform throughout the ejecta.
(The black dashed line in each panel corresponds to a perfectly mixed outflow.)
However, as was true for \Mw{} (\S\ref{subsec:mw_ew}), the impact of \tauw{} decreases with \Ei.
For example, for wind-driven models, increasing \tauw{} from $0.1R_0/c$ to $10R_0/c$ changes the mixing profile from one in which the wind material resides mainly in a central core to one in which the wind has mixed almost evenly into the ejecta.
For models with $\Ei/M_0c^2 = 5 \times 10^{-3}$, the highest \tauw{} is still associated with the most mixing, but the contrast with lower-\tauw{} cases is much less stark; even the best-mixed model concentrates the wind material in the ejecta's center.

The greater degree of mixing and the somewhat bumpier mixing profiles that characterize models with higher \tauw{} are both attributable to Raleigh-Taylor instabilities at the wind-ejecta interface, which have more time to develop when \tauw{} is longer. 
For sufficiently long \tauw, these instabilities produce more spatially extended plumes that are less easily smoothed by angle-averaging.
This can be seen in Fig.~\ref{fig:tauw_eddies}, which shows the final wind mass fraction as a function of velocity coordinate and polar angle for the wind-driven models of Fig.~\ref{fig:tau_fx}.
The instabilities that develop for $\tauw \gtrsim R_0/c$ transport a significant fraction of the wind mass to higher velocities and mass coordinates, producing the more homogeneous  mixing profiles apparent in Fig.~\ref{fig:tau_fx}.

\begin{figure}\includegraphics[width=\columnwidth]{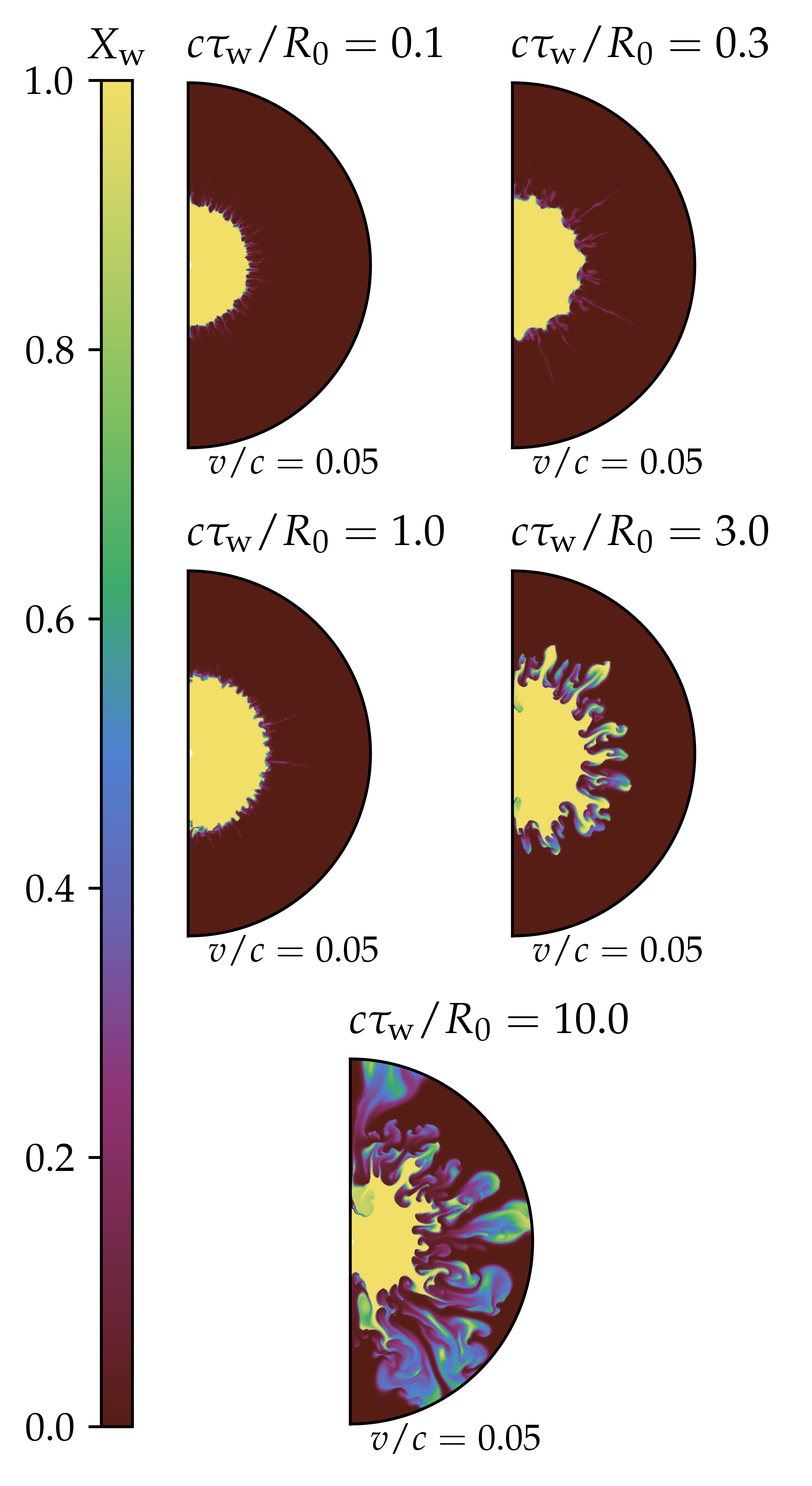}
    \caption{Longer wind durations (\tauw) more effectively distribute wind material to higher velocity and mass coordinates, and produce large-scale structures originating in the wind-ejecta boundary.
    Each panel above shows the final wind mass fraction \Xw{} as a function of the radial velocity coordinate and the polar angle for one of the wind-driven models of Fig.~\ref{fig:tau_fx} ($\Mw = 0.1M_0$, $\tiw = 10^{-3}R_0/c$).
    For $\tauw \leq R_0/c$, the wind material is mostly confined to a central core. 
    For longer \tauw{}, instabilities develop, allowing tendrils of wind matter to punch through to the outer layers of the ejecta.}
    \label{fig:tauw_eddies}
\end{figure}

We quantify the sensitivity of \Xw{} to \tauw{} by modifying the $\zeta$ parameter introduced in Eq.~\ref{eq:zeta_ME}.
Specifically, the dependence on wind duration can be represented by a power law, yielding an updated definition,

\begin{align}
    \zeta(\Mw,\Ei,\tauw) = \zeta(\Mw,\Ei)\; \left(\frac{
    \tauw}{R_0/c}\right)^{3/4}.\label{eq:zeta_MET}   
\end{align}

Eq.~\ref{eq:zeta_MET} explains the variation in mixing patterns for all models considered thus far, as can be seen in Fig.~\ref{fig:zeta_met}.
The top panel presents the wind mixing profile for every model of \S\ref{subsec:mw_ew} and \S\ref{subsec:tau_wind}, color-coded by $\zeta(\Mw,\Ei,\tauw)$.
We also characterize each mixing profile using a single parameter $\mbar_{95}$, defined as the value of the modified mass coordinate \mbar{} within which 95\% of the wind mass is contained.
We plot this quantity as a function of $\zeta(\Mw,\Ei,\tauw)$ in Fig.~\ref{fig:zeta_met}'s lower panel.\footnote[2]{The model $(\Mw, \Ei, \tauw) = (0.5M_0,\; 5 \times 10^{-3}M_0c^2, \; R_0/c)$ has $\mbar_{95} = -0.01$ (the 95\% threshold means values below zero are technically possible) and cannot be plotted on a log scale, so is omitted from the lower panel.}
Though some scatter is evident, $\zeta(\Mw, \Ei, \tauw)$ is clearly predictive of mixing.

\begin{figure}\includegraphics[width=\columnwidth]{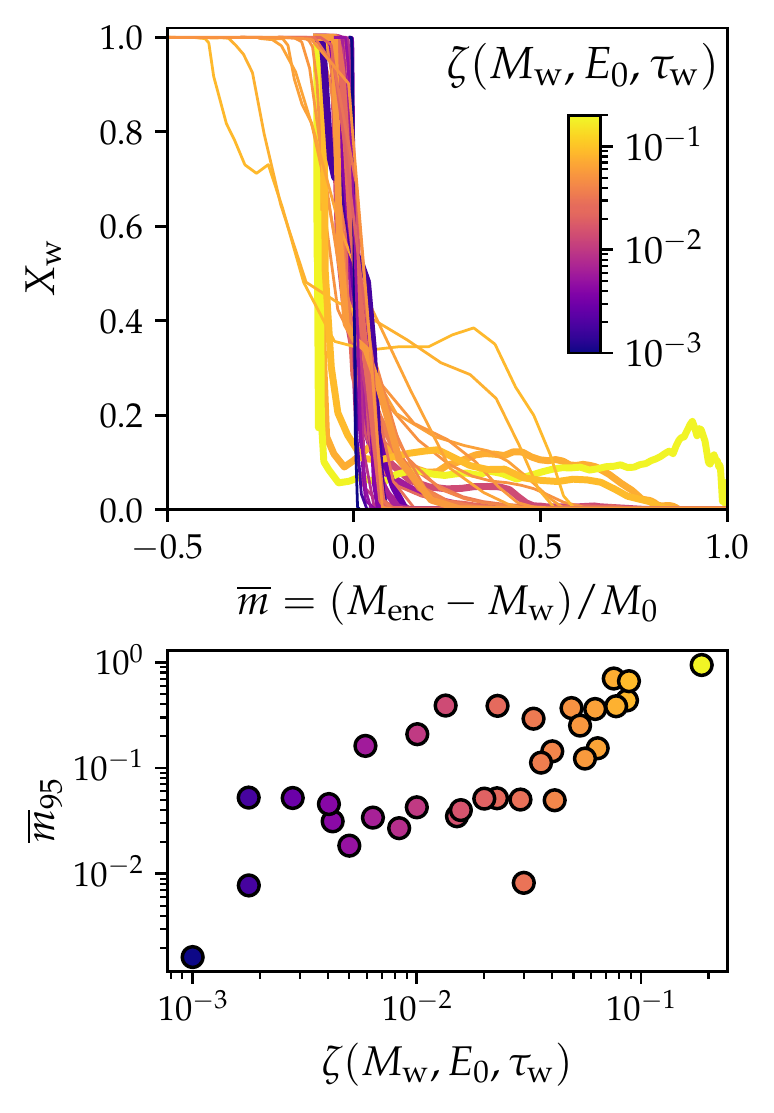}
\caption{The wind duration \tauw{} influences mixing between the wind and the ejecta.
\emph{Top panel:} The wind mixing profiles for all models of \S\ref{subsec:mw_ew} and \S\ref{subsec:tau_wind}, color-coded by the redefined $\zeta$ parameter (Eq.~\ref{eq:zeta_MET}) which now accounts for the effects of \tauw.
The profiles for models with $\tauw \neq R_0/c$ ($\tauw = R_0/c$) are plotted as thick (thin) lines, to make it easier to discern which curves have been added since Fig.~\ref{fig:mix_by_zeta}.
\emph{Bottom panel:} We calculate for each  model $\mbar_{95}$, the modified  mass coordinate (\mbar) inside  which 95\% of that model's wind mass is contained, and plot it versus $\zeta(\Mw, \Ei, \tauw)$ to validate the updated expression for the latter.
The essentially linear relationship between $\zeta$ and $\mbar_{95}$ increases confidence in Eq.~\ref{eq:zeta_MET}.
}\label{fig:zeta_met}
\end{figure}

\subsection{Effect of wind start time}\label{subsec:wind_t0}

The calculations of \S\ref{subsec:mw_ew} and \S\ref{subsec:tau_wind} all assume a disk wind whose launch coincides with an initial explosion, if there is one. 
However, if it takes time for the infalling material to circularize around the central remnant \citep[e.g.][]{MacFadyen.ea_2001.ApJ_collapsars.jets.sne,Dessart.eq_2008.ApJ_PNS.collapsars}, a precondition for the formation of the disk that eventually generates the wind, the wind could conceivably lag any prompt explosion.

To understand how a delay in the wind launch affects mixing, we introduce a wind start time parameter, \tiw.
We explore the impact of \tiw{} using a subset of models with $(\Mw, \Ei, \tauw) = (0.1M_0, \; 4.5 \times 10^{-4}M_0c^2, \; R_0/c)$.
In addition to the default value ($\tiw = 10^{-3}R_0/c$), we consider $\tiw = \beta R_0/c$, with $\beta = $ 0.1, 0.3, 1, 3, and 10.

In conceptualizing these values, it is instructive to compare the wind start time to the free-fall time $t_{\rm ff}$ of the progenitor star,

\begin{align}
\frac{t_{0, \rm w}}{t_{\rm ff}} = 1.3 \times 10^{-3} \; \beta \; \left(\frac{M_0/\msun}{R_0/R_\odot}\right)^{1/2}.\label{eq:tff_ratio} 
\end{align}
As Eq.~\ref{eq:tff_ratio} makes clear, if $M \sim \msun$ and $R \sim R_\odot$, as we can reasonably expect for the progenitors of GRB-SNe and SNe Ic-BL (see \S\ref{subsec:tau_wind}), even our longest $\tiw$ accounts for barely 1\% of $t_{\rm ff}$.
In other words, the wind is launched (and therefore the disk is assumed to form) on timescales short relative to the free-fall time.
This is appropriate, since disk formation is enabled by the inner stellar layers, which have a free-fall/circularization time much lower than that of the star as a whole.
\citet{MacFadyen.ea_2001.ApJ_collapsars.jets.sne} find a similarly low ratio of accretion disk formation time to free-fall timescale, although their progenitor star is more massive and radially extended than ours.

\begin{figure}\includegraphics[width=\columnwidth]{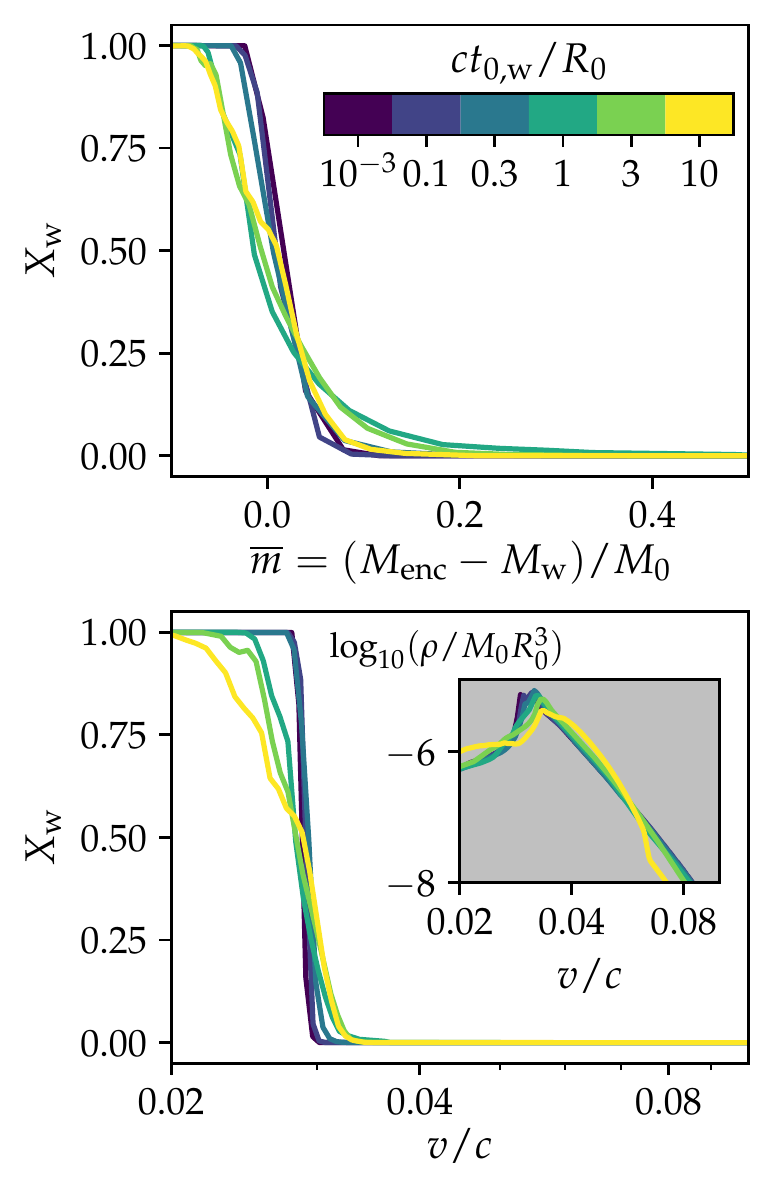}
    \caption{Delaying the wind, relative to the initial explosion, increases mixing, but the effect is minor. The models plotted above differ only in \tiw, and all have $(\Mw, \Ei, \tauw) = (0.1 M_0, \; 4.5 \times 10^{-4} M_0 c^2, \;R_0/c)$.
    \emph{Top panel:} the wind mass fraction, \Xw, is shown as a function of \mbar, as in Figs.~\ref{fig:xmix_ME} and \ref{fig:tau_fx}.
    Mixing generally increases with \tiw, but the relationship is not strictly monotonic.
    \emph{Bottom panel:} \Xw{} as a function of velocity, which more clearly reveals the connection between mixing and the wind start time.
    \emph{Inset:} Final density profiles for each model. 
    The density profile for $\tiw=10 R_0/c$ (yellow curve) is distinct from those of the other models, which obscures the relationship between \tiw{} and $\Xw(\mbar)$.
    This may be because the extreme delay in the $\tiw=10R_0/c$ case pushes the explosion into a different hydrodynamic regime.
    (See text for a full discussion.)
    }
    \label{fig:t0w_mix}
\end{figure}

Increasing \tiw{} increases mixing, but the effect is weak, as demonstrated by Fig.~\ref{fig:t0w_mix}.
The top panel shows the final wind mass fraction, \Xw, as a function of \mbar.
While a correlation between long delays and enhanced mixing is evident, the pattern is not perfectly monotonic, owing to the anomalously low mixing in the model with $\tiw=10R_0/c$.

In the lower panel, we plot \Xw{} as a function of velocity coordinate, rather than \mbar, and recover the expected monotonicity.
The reason underlying the discrepancy can be understood from the inset in the lower panel, which shows the final density profile for each of the six models with variable \tiw.
The shape of the profile corresponding  to $\tiw = 10R_0/c$ has a broader density peak, and an elevated density inside that peak, compared to models with lower \tiw.
As a result, its enclosed mass evolves with velocity in a manner distinct from that which characterizes the models with $\tiw \leq 3 R_0/c$.
In this case, plotting quantities of interest with respect to \mbar{} can  obscure the underlying relationships.

The unique behavior of $\rho(v)$ for $\tiw = 10 R_0/c$ most likely indicates that this delay is long enough to push the system into a new hydrodynamic regime characterized by the interaction of thin shells, rather than spatially extended outflows.
Given the insensitivity of mixing to \tiw, even when the latter is varied by four orders of magnitude, we do not further investigate this limiting behavior here.

\section{Implications for Supernova Emission}
\label{sec:sn_emission}
Though we do not undertake an exhaustive survey of the emission that would result from all the explosion models we study here, to elucidate general trends we present synthetic photometry for a handful of cases and discuss the implications of our findings.

\subsection{Model subset photometry}\label{subsec:lcs_tauw}
While the \rp{} mixing pattern is known to impact the emission of the associated SN \citep{Siegel.Barnes.Metzger_2019.Nature_rp.collapsar}, \citetalias{Barnes.Metzger_2022.ApJL_rproc.collapsar} demonstrated that the nature of the impact depends additionally on factors such as the total ejected mass and the kinetic energy of the SN. 

In light of these overlapping dependencies, models with fixed \Ei{}, fixed $\Mw$, and variable wind durations (\S\ref{subsec:tau_wind}) represent ideal test cases for exploring the effects of the various mixing patterns apparent in the full model suite.
While they evince a range of mixing behaviors (Fig.~\ref{fig:tau_fx}), their shared properties enable a more apples-to-apples comparison than would be possible for other model groupings. 

We focus on models with $\Mw = 0.1M_0$, $\Ei = 4.5 \times 10^{-4} M_0 c^2$, and $c\tauw/R_0 = 0.1$, $1$, $3$, and $10$.
These models have $\zeta(\Mw,\Ei,\tauw) = 3\times 10^{-3}$, $1.6 \times 10^{-2}$, $3.6 \times 10^{-2}$, and $8.8 \times 10^{-2}$, respectively. 
As Eq.~\ref{eq:zeta_MET} and Fig.~\ref{fig:zeta_met} indicate, similar mixing patterns can be achieved by other parameter combinations. 

To calculate the SN emission, we first angle-average the final mass density profile for each model, and scale it so all models have a total mass, including the wind component, of $\mej = M_0 + \Mw = 4\msun$, a value typical of SNe Ic-BL \citep[e.g.,][]{Taddia.ea_2019.AandA_IcBL.iPTF.survey}.
(If we assume a remnant black hole mass $M_{\rm BH} \gtrsim 1.4 \msun$, this requires a modest rescaling of our progenitor density profile.)
With this choice,
each model represents a SN with a wind (non-wind ejecta) mass of $\Mw = 0.36\msun$ ($M_0 = 3.64\msun$, and a final kinetic energy $\ekin = 6.2 \times 10^{51}$ erg.
The characteristic velocity is $0.04c$, lower than the fastest expanding SNe Ic-BL, but not inconsistent with the general population \citep{Modjaz_etal_2016_GrbSne,Taddia.ea_2019.AandA_IcBL.iPTF.survey}.
The assumption of homologous expansion (valid on timescales relevant to the light curve) completely specifies the density profile as a function of time.

We likewise angle-average \Xw{}.
Our SN models also include \nickel, the synthesis and mixing of which is not calculated by \texttt{JET}.
Instead, we assume that \nickel{} is evenly distributed in the non-wind SN ejecta, so 

\begin{equation}
X_{56}(v) = \frac{M_{56}}{M_0}(1-\Xw),\label{eq:X56}
\end{equation}
with $X_{56}$ the \nickel{} mass fraction and \mni{} the total mass of \nickel.
We adopt $\mni = 0.25\msun$, in line with expectations of SNe with our specified \mej{} and \ekin{} \citep{Taddia.ea_2019.AandA_IcBL.iPTF.survey}.

With this assumption, we have all the ingredients required to perform radiation transport  using the semianalytic framework introduced in \citetalias{Barnes.Metzger_2022.ApJL_rproc.collapsar}.
As in that work, we assume that \rp{} material carries an opacity $\kappa_{r \rm p} = 10$ cm$^2$ \per{g}, the opacity of \nickel{} is temperature dependent, and unenriched SN ejecta has an opacity $\kappa_{\rm sn} = 0.05$ cm$^2$ \per{g}. 
The spectrum of optically thin \emph{r}-process material is taken to be well-approximated by a blackbody of temperature $T = 2500$ K (see \citet{Hotokezaka.ea_2021.MNRAS_kn.nebular.model} or \citetalias{Barnes.Metzger_2022.ApJL_rproc.collapsar} for details).

The broadband light curves for the four models are presented in the top panels of Fig.~\ref{fig:lc_bb_comp}.
To better disentangle the effects of \rp{} enrichment from those due to other factors (changes to the density profile, for example), we show in the lower panels a variant of each model that assumes the wind material has an opacity equal to that of ordinary SN ejecta.
(Put another way, these models correspond to an explosion scenario in which a wind is launched, but that wind is free of both \rp{} elements and \nickel). 

\begin{figure*}\includegraphics[width=\textwidth]{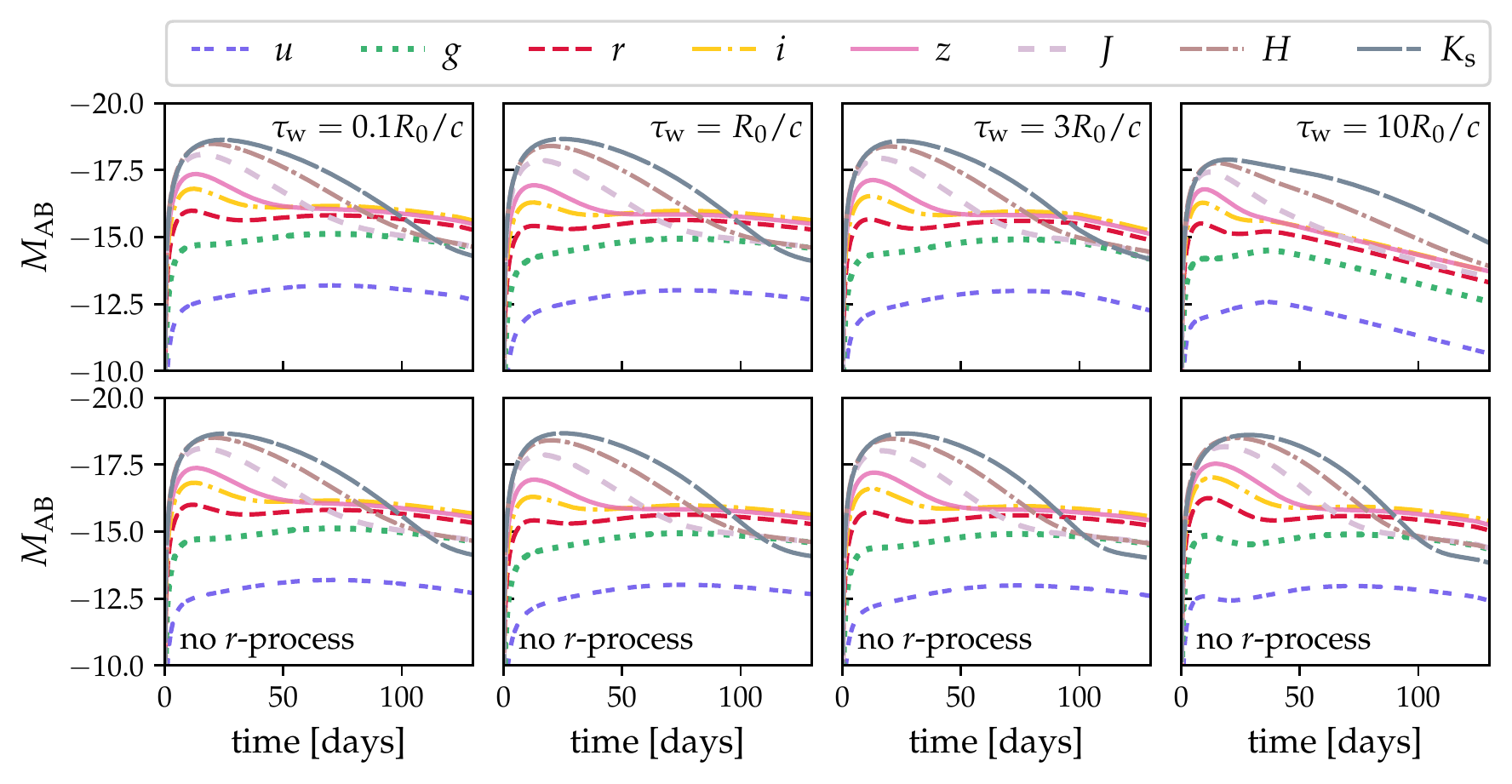}
\caption{The impact of \rp{} material on SN emission depends on its distribution within the ejecta. 
We show broadband light curves calculated using a semianalytic method for four models that share a wind mass, total mass, \nickel{} mass, and kinetic energy (equal to $0.36\msun$, $4.0\msun$, $0.25\msun$, and $6.2 \times 10^{51}$ erg, respectively), but have different levels of mixing due to their differing wind durations. 
The upper panels show light curves assuming the wind is composed purely of \rp{} elements.
For comparison, we show in the lower panels a variation on each model, which assumes the wind material is non-radioactive and has the same opacity as ordinary \nickel-free SN ejecta ($\kappa = 0.05$ cm$^2$ \per{g}).
The light curves of the \rp-enriched models resemble those of the \rp-free variants for $\tauw \leq 3R_0/c$, corresponding to a mixing parameter $\zeta \leq 0.036$.}\label{fig:lc_bb_comp}
\end{figure*}

The \rp-enriched models of the top panels are similar to each other---and similar to their \rp-free counterparts---for $0.1 \leq c\tauw/R_0 \leq 3$ ($5 \times 10^{-3} \leq \zeta \leq 3.6 \times 10^{-2}$), demonstrating that at lower mixing levels, the effect of \rp{} enrichment is minimal, at least for this \mej{} and \ekin.
The situation changes for $\tauw = 10 R_0/c$.
Compared to the other \rp-enriched models of Fig.~\ref{fig:lc_bb_comp}'s top row, this case shows significant late-time brightening in the near infrared (NIR) $J$, $H$, and $K_{\rm s}$ bands at the expense of optical emission.
The resemblance of the \rp-free model with $\tau=10R_0/c$ to the \rp-free models with $\tauw \leq 3R_0/c$ implies that the principal driver of these effects is the distribution of \rp{} material within the ejecta, rather than differences in the density profile resulting from the longer wind duration. 

To further clarify the effects of mixing, we show in Fig.~\ref{fig:deltaRX} the difference in the $r-K_{\rm s}$ color, $\Delta(r - K_{\rm s})$, between the \rp-enriched and \rp-free models of Fig.~\ref{fig:lc_bb_comp}.
For $\Delta(r - K_{\rm s}) = 0$, the NIR flux of a given \rp-enriched model, relative to its optical flux, is identical to that of its unenriched analogue, while  $\Delta(r - K_{\rm s}) < 0$ indicates enhanced NIR emission from the enriched SN.

\begin{figure}\includegraphics[width=\columnwidth]{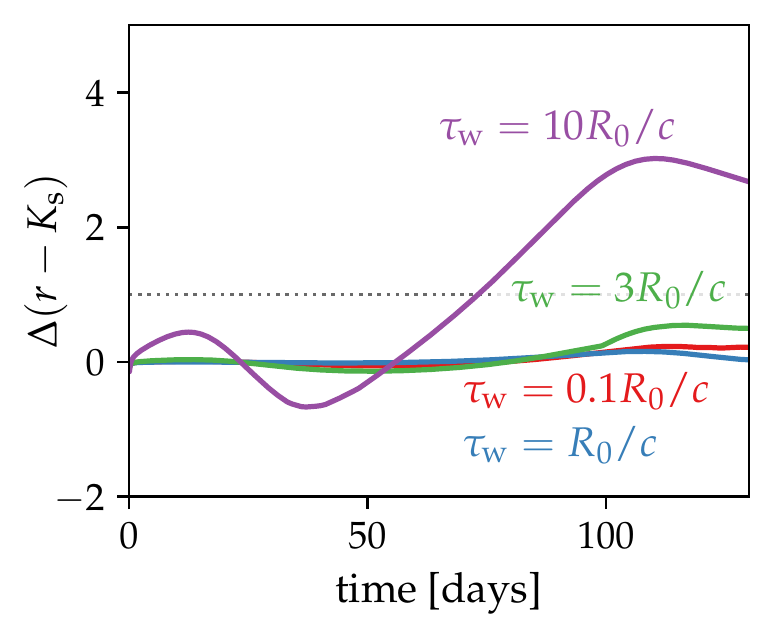}
    \caption{Extensive mixing enhances late-time NIR emission at the expense of optical flux.
    We show how the predicted $r-K_{\rm s}$ color changes for models with  $M_0 = 3.64\msun$, $\Mw = 0.36\msun$, $\ekin = 6.2 \times 10^{51}$ erg, and various values of $\tauw$ if the disk wind is composed of \rp{} elements, rather than ordinary stellar material. 
    (In other words, we show the $r-K_{\rm s}$ color differences between the models in the top and bottom rows of Fig.~\ref{fig:lc_bb_comp}.)
    The impact on color is minor except for $\tauw = 10R_0/c$ ($\zeta = 0.088$).}
    \label{fig:deltaRX}
\end{figure}

When mixing is negligible, as it is for $c\tauw/R_0 = 0.1$ and $1.0$, the difference in $r-K_{\rm s}$ due to the \rp{} material is effectively undetectable.
At slightly higher levels of mixing (e.g., $\zeta = 0.036$ for $\tauw = 3R_0/c$), the color difference is ${\lesssim}0.5$ mag, and appears only for $t \gtrsim 100$ days.
It is only for our best mixed model, with $\tauw = 10 R_0/c$ and $\zeta = 0.088$, that the \rp{} material impacts the colors significantly, and even here, $\Delta(r - K_{\rm s})$ does not exceed 1 mag until ${\sim}75$ days after the explosion.
It appears that, given the particular density profiles that arise from our wind-augmented explosion models, substantial mixing is required to alter the SN light curves.

Juxtapositions like those of Figs.~\ref{fig:lc_bb_comp} and \ref{fig:deltaRX} are useful for demonstrating the relationship between mixing and reddening---namely, that if other parameters are equal, more mixing equates to more reddening at earlier times.
However, we advise against reading too much into these comparisons, as the \rp-free reference cases were selected assuming a particular explosion scenario (an initial spherical explosion enhanced by an \rp-free disk wind) that may not be realized in nature.
Furthermore, the specifics of the \rp{} signal evident in Figs.~\ref{fig:lc_bb_comp} and \ref{fig:deltaRX} are a function of our chosen \mej{}, \mni{}, and \ekin, as well as of the \rp{} parameters.
(More detailed discussions of how to make appropriate comparisons between \rp-enriched and \rp-free models can be found in \citet{Barnes.Metzger_2022.ApJL_rproc.collapsar} and \citet{Anand.ea_2023.arXiv_rp.collapsar.survey}.)

Thus, while this analysis is useful as a proof of concept, the conclusions we draw from it cannot be trivially extended to other regions of parameter space.
Bearing this caveat in mind, we cautiously embark on a final exploration of the corruption of SN signals by well-mixed \rp{}-wind material. 

\subsection{Wind-dominated supernovae}\label{subsec:windy-sn}
We now turn our focus to wind-dominated SNe---explosions that derive most of their kinetic energy from disk outflows.
Because the standard neutrino mechanism \citep{ColgateWhite_1966_NeutrinoMech, BetheWilson_1985_NeutrinoMech} cannot account for the high values of \ekin{} ascribed to SNe Ic-BL \citep[e.g.,][]{Iwamoto.ea_1998.Natur_icbl.model.1998bw,Mazzali2014_Magnetars}, alternate energy sources must be invoked to explain these events. 
Disk winds have been a favored engine candidate since the inception of the collapsar model \citep{MacFadyenWoosley99_collapsar}.

We consider SNe of total mass $\mej = 4.0\msun$, as in \S\ref{subsec:lcs_tauw}, but slightly higher $\ekin = 8 \times 10^{51}$ erg, because we are interested in the SNe whose energies most strongly suggest an unusual explosion mechanism.
In a departure from earlier sections, we now treat the wind velocity as a free parameter, rather than fixing $\vw = 0.1c$.
Specifying \Ei{} and \Mw{} thus determines $\vw$ through the requirement that the initial explosion energy and wind energy sum to \ekin.

Together with the wind duration \tauw, these parameters determine the value of $\zeta$ (Eq.~\ref{eq:zeta_MET}), which quantifies the extent of the mixing.
For standard SNe Ic-BL masses and velocities, the arguments of \S\ref{subsec:lcs_tauw} (Figs.~\ref{fig:lc_bb_comp} and \ref{fig:deltaRX}) identify $\zeta \approx 0.09$ as a rough threshold separating \rccsne{}  with a detectable \rp{} signal from those whose \rp{} enrichment is well hidden.
In Fig.~\ref{fig:mt_crit}, we show for a handful of initial explosion energies the combination of wind masses and durations that produce $\zeta = 0.09$ for SNe with $\mej$ $(\ekin)$ equal to $4.0\msun$ ($8\times 10^{51}$ erg).

\begin{figure}\includegraphics[width=\columnwidth]{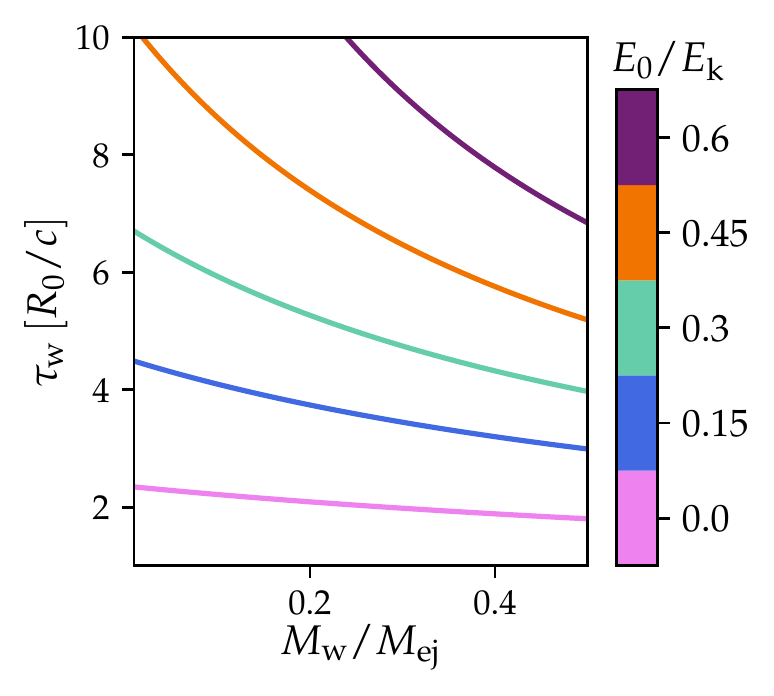}
    \caption{Whether the presence of \rp{} material in the ejecta can be easily diagnosed from SN emission depends on a combination of parameters. 
    We calculate the \Mw{} and \tauw{} that yield $\zeta = 0.09$ (a value that indicates extensive mixing and adulteration of the SN signal) for a SN with $\mej = 4.0\msun$ and $\ekin = 8 \times 10^{51}$ erg, for various initial explosion energies \Ei.
    In contrast to earlier sections, we now allow the wind velocity to vary.
    Of particular interest are wind-driven SNe ($\Ei = 0$; pink curve).
    To achieve the specified \ekin, wind-driven models must have massive and/or high-velocity winds that cause high levels of mixing even at fairly low \tauw.
    }
    \label{fig:mt_crit}
\end{figure}

Fig.~\ref{fig:mt_crit} reflects some of the trends identified earlier.
For example, for a given \Mw{}, $\zeta$ decreases with \Ei, so longer \tauw{} are required to achieve the same level of mixing. 
However, relaxing the assumption that $\vw = 0.1c$ also enables new inferences.

In particular, Fig.~\ref{fig:mt_crit} shows that if explosions with typical SNe Ic-BL masses and kinetic energies are powered exclusively by a disk wind ($\Ei = 0$; pink curve), and that disk wind is composed of freshly synthesized \rp{} elements, the mixing will be extensive enough to impact the SNe signal as long as $\tauw \gtrsim 2R_0/c$, regardless of the wind mass.
The stripped-envelope stellar progenitors believed to end their lives as SNe Ic-BL have pre-explosion radii comparable to $R_\odot$; for such stars, $2R_0/c$ corresponds to a wind timescale of $1 \text{ s } \lesssim \tauw \lesssim 10$ s.

If we assume that GRB durations are rough indicators of accretion disk (and therefore wind) lifetimes, we conclude that \tauw{} comfortably exceeds the critical value in all but the shortest-duration long GRBs. 
This implies either that typical SNe Ic-BL derive their energy from a source other than (or in addition to) a disk wind, or that the disk winds that power SNe Ic-BL are not composed of heavy \rp{} material with uniquely high opacities.\footnote[3]{An additional question that arises for wind-driven \rccsne{} is the origin of \nickel, which for $\Ei=0$ cannot be attributed to a prompt explosion. While nucleosynthesis is not our current focus, we note that some authors \citep[e.g.,][]{Maeda.Tominaga_2009.MNRAS_nickel.wind.driven.sne} have found that \nickel-burning may occur in stellar gas outside the disk when it is shocked by outflowing winds.} 

\section{Conclusion}\label{sec:conclusion}

We used two-dimensional hydrodynamic simulations to study mixing processes in SNe that feature a disk wind either as their exclusive kinetic energy source or in addition to an initial explosion. 
If \rp-rich disk outflows are a near-universal component of GRB-SNe and/or SNe Ic-BL, as proposed by the \crp{} theory, characterizing disk-ejecta mixing is essential for understanding the impact of \rp{} pollution on SN signals.

We modeled wind-enhanced SN explosions using four parameters, and discovered that mixing increases with wind mass (\Mw) and duration (\tauw)  and decreases with the initial explosion energy (\Ei).
We derived a straightforward and physically motivated way to quantify these dependencies (Eq.~\ref{eq:zeta_MET}).
Intriguingly, the sensitivity of mixing to \tauw{} decouples mixing from the wind mass and energy; a range of outcomes is possible at any given \Mw{} and \ekin.
The effect of the fourth parameter, \tiw{} (the start time of the wind relative to the initial explosion), is minor compared to the impact of the other three variables.

Having simulated the explosions of \rccsne{} with disk winds, we next used semianalytic radiation transport to understand how the density and mixing profiles of these SNe affect their electromagnetic emission. 
We focused on SNe with ejecta masses, velocities, and 
\nickel{} masses typical of SNe Ic-BL, and discovered that only fairly high levels of mixing appreciably altered the emission relative to cases with \rp-free winds. 
Given that one way to achieve this degree of mixing is through an extended wind duration, this finding identifies SNe associated with longer long GRBs as ideal targets in the search for \rccsne.

Extrapolating the trends we observed in our survey of the $\Mw$-$\Ei$-$\tauw$ parameter space allowed us to argue that typical SNe Ic-BL are not likely to be both wind-driven and \rp{} enriched.
If they were, their emission would exhibit reddening inconsistent with observations. 
Thus, if the \crp{}  hypothesis holds, we cannot appeal to high-velocity disk winds to account for the anomalously high \ekin{} inferred for GRB-SNe and SNe Ic-BL.
On the other hand, if \nickel{} is synthesized in the wind \citep{Pruet.ea_2003.ApJ_nucleosyn.grb.disks,Nagataki.ea_2006.ApJ_hypernova.nucleosyn,Maeda.Tominaga_2009.MNRAS_nickel.wind.driven.sne}, a wind-driven explosion could naturally explain the extensive \nickel-mixing inferred for SNe Ic-BL \citep{Taddia.ea_2019.AandA_IcBL.iPTF.survey}.

While this work has uncovered important relationships between explosion parameters, wind-ejecta mixing, and SN emission, uncertainties remain. 
In particular, the SN signal depends on the distribution of \nickel{} within the ejecta, a sensitivity we neglected here.
Future work should incorporate nuclear network calculations to model the production and subsequent mixing of \nickel{} in the SN ejecta.
Additional avenues for exploration include the role of the GRB jet in hydrodynamic mixing and the effect of the progenitor structure on the trends we discovered. 
The latter may be of particular interest as recent observational evidence \citep{Taddia.ea_2019.AandA_IcBL.iPTF.survey} suggests that both single and binary stripped stars can evolve to GRB SNe and SNe Ic-BL. 
These two classes can have very different pre-explosion radii \citep[e.g.][]{Laplace.ea_2021.A&A_stripped.stars.single.binary,Schneider.ea_2021A&A_stripped.stars.single.binary}, which necessarily impacts the mapping from GRB durations to $c\tauw/R_0$.

The site(s) of \rp{} production remains a topic of active investigation.
Our understanding of astrophysical heavy-element production is likely to expand in the near future as the LIGO-Virgo-KAGRA gravitational-wave detector network identifies new NSMs and as new instruments (e.g., NIRSpec aboard \emph{JWST}) provide unprecedented opportunities to probe the reddest emission from all candidate \rp{} production sites.
Constructing realistic models of \rp{} mixing in \rccsne{} is critical for leveraging these capabilities, assessing the \crp{} hypothesis, and unravelling the question of \rp{} origins.

\vspace{\baselineskip}
The authors thank the anonymous referee for comments that helped us improve the manuscript, as well as J. Fuller for helpful discussions.
J.B. gratefully acknowledges support from the Gordon and Betty Moore Foundation through grant No. GBMF5076.
Additional support was provided by the National Science Foundation under grant No. NSF PHY-1748958.
P.D. acknowledges support from the National Science Foundation under grant No. NSF AAG-2206299, and by NASA under grant No. 21-FERMI21-0034.

\bibliographystyle{apj} 
\bibliography{refs}

\end{document}